\documentclass[10pt,preprint,twocolumn]{aastex7}
\newcommand{\roma}[1]{\uppercase\expandafter{\romannumeral#1}}

\usepackage{amsmath}
\usepackage{subfigure}
\usepackage{hyperref}

\usepackage{lineno}
\usepackage{color}
\usepackage{makecell}
\usepackage{graphicx}
\usepackage{natbib}
\usepackage{amssymb,txfonts}
\usepackage{multirow}
\usepackage{array}
\usepackage{sidecap}
\usepackage{hyperref}
\usepackage{epstopdf}
\usepackage{footnote}
\usepackage{tabularx}
\usepackage{booktabs}

\shorttitle{An observational support of the cancellation nanoflare model}
\shortauthors{Tang et al.}
\received{February 12, 2025}
\revised{March 26, 2025}
\accepted{\today}
\submitjournal{APJL}

\begin{document}
\title{High-Resolution Observations of a Small-Scale Cancellation Nanoflare: Supporting Evidence for the Cancellation Nanoflare Model}
\correspondingauthor{Yuandeng Shen}

\author{Zehao Tang$^{1,4}$}
\noaffiliation{}
\affiliation{Yunnan Observatories, Chinese Academy of Sciences, Kunming 650216, China}
\affiliation{State Key Laboratory of Solar Activity and Space Weather, School of Aerospace, Harbin Institute of Technology, Shenzhen 518055, China}
\affiliation{Shenzhen Key Laboratory of Numerical Prediction for Space Storm, Harbin Institute of Technology, Shenzhen 518055, China}
\affiliation{University of Chinese Academy of Sciences, Beijing, 100049, China}
\email{tangzh@ynao.ac.cn}  

\author[orcid=0000-0001-9493-4418]{Yuandeng Shen}
\affiliation{State Key Laboratory of Solar Activity and Space Weather, School of Aerospace, Harbin Institute of Technology, Shenzhen 518055, China}
\affiliation{Shenzhen Key Laboratory of Numerical Prediction for Space Storm, Harbin Institute of Technology, Shenzhen 518055, China}
\email[show]{ydshen@hit.edu.cn}

\author{Chengrui Zhou}
\affiliation{Yunnan Observatories, Chinese Academy of Sciences, Kunming 650216, China}
\affiliation{University of Chinese Academy of Sciences, Beijing, 100049, China}
\email{zhouchengrui@ynao.ac.cn}

\author{Surui Yao}
\affiliation{Yunnan Observatories, Chinese Academy of Sciences, Kunming 650216, China}
\affiliation{University of Chinese Academy of Sciences, Beijing, 100049, China}
\email{yaosurui@ynao.ac.cn}

\author{Dongxu Liu}
\affiliation{Yunnan Observatories, Chinese Academy of Sciences, Kunming 650216, China}
\affiliation{University of Chinese Academy of Sciences, Beijing, 100049, China}
\email{liudongxu@ynao.ac.cn}

\begin{abstract}
An analytical cancellation nanoflare model has recently been established to show the fundamental role that ubiquitous small-scale cancellation nanoflares play in solar atmospheric heating. Although this model is well-supported by simulations, observational evidence is needed to deepen our understanding of cancellation nanoflares. We present observations of a small-scale cancellation nanoflare event, analyzing its magnetic topology evolution, triggers, and physical parameters. Using coordinated observations from Solar Dynamics Observatory and Goode Solar Telescope, we identify a photospheric flow-driven cancellation event with a flux cancellation rate of $\sim10^{15}$ Mx s$^{-1}$ and a heating rate of $8.7 \times 10^6$ erg cm$^{-2}$ s$^{-1}$. The event shows the characteristic transition from $\pi$-shaped to X-shaped magnetic configuration before forming a 2$\arcsec$ current sheet, closely matching model predictions. This event provides critical observational support for the cancellation nanoflare model and its role in solar atmospheric heating.
\end{abstract}
\keywords{magnetic reconnection --- Sun: activity --- Sun: magnetic fields --- Sun: chromosphere --- magnetohydrodynamics (MHD)}

\section{Introduction}
The temperature structure of the solar atmosphere presents one of the most intriguing puzzles in solar physics. While temperature initially decreases with height, reaching a minimum of ~4,200 K at the top of the photosphere, it exhibits an unexpected reversal. The temperature rises gradually to ~20,000 K through the chromosphere, followed by a dramatic increase through the narrow transition region (reaching several 100,000 K) to approximately 2 MK in the corona. This peculiar temperature profile, seemingly defying basic thermodynamic principles, constitutes the long-standing solar atmospheric heating problem \citep{2004psci.book.....A}. A particularly striking aspect is that the energy required for coronal heating represents merely $10^{-4}$ of the Sun's total energy output. Thus, the central question shifts from identifying the energy source to understanding the mechanisms responsible for its transport and deposition in the solar atmosphere.

Magnetic fields have emerged as the key to unraveling this mystery. Generated in the solar interior through dynamo processes, these fields emerge into the atmosphere via magnetic buoyancy. However, their atmospheric presence is inherently transient. The fields inevitably interact with opposite-polarity regions in a process known as magnetic cancellation \citep{1985AuJPh..38..855L, 1985AuJPh..38..929M}, which serves as the primary mechanism for removing magnetic flux from the solar atmosphere. While \citet{1985SoPh..100..397Z, 1987ARA&A..25...83Z} initially proposed that magnetic flux cancellation could occur through either submergence or magnetic reconnection, subsequent observations have strongly favored the reconnection scenario. Key evidence includes the behavior of same-polarity features, which merge rather than cancel upon interaction, and the prevalence of cancellation events in the presence of external fields \citep{1999SoPh..190...35H, 1999ApJ...515..435L, 2017ApJ...845...94T, 2018NewA...65....7T, 2021ApJ...912L..15T, 2021ApJ...923...45Z}.

The significance of magnetic cancellation extends far beyond simple flux removal. These interactions are crucial drivers of various solar phenomena across multiple scales. At smaller scales, cancellation events trigger spicules \citep{2019Sci...366..890S}, Ellerman bombs \citep{2016ApJ...823..110R}, and ultraviolet bursts \citep{2014Sci...346C.315P}. On larger scales, they play essential roles in solar jet formation \citep{2011ApJ...735L..43S,  2012ApJ...745..164S, 2016ApJ...832L...7P, 2018ApJ...864...68S, 2021RSPSA.47700217S}, filament development \citep{2024ApJ...975L...5Y}, and potentially contribute to solar wind generation. This multi-scale influence is particularly evident in quiet Sun regions, where network magnetic fields maintain a dynamic equilibrium through the continuous interplay of flux emergence and cancellation \citep{2018ApJ...857...48G, 2004Natur.430..326T}.

The complex dynamics of these magnetic interactions arise from their anchoring in the solar photosphere. In this high-$\beta$ plasma environment, magnetic field footpoints are continuously displaced by granular motions and advection. This constant buffeting leads to a highly structured magnetic field distribution, with fields preferentially concentrating along intergranular lanes and forming the characteristic magnetic network in quiet Sun regions. Early observations estimated a quiet-Sun cancellation rate of $10^{14}$ Mx s$^{-1}$ \citep{1985AuJPh..38..855L}. However, recent high-resolution studies by \citet{2017ApJS..229...17S} have revealed that small-scale cancellations occur far more frequently than previously thought, with a combined cancellation rate reaching $10^{15}$ Mx s$^{-1}$. This higher rate suggests these events may constitute a significant energy source for atmospheric heating \citep{1998SoPh..178..245Z}.

The relationship between magnetic cancellation and atmospheric heating has been extensively investigated through observational and theoretical approaches. Observational studies have consistently found atmospheric local heating at cancellation sites \citep{2007ApJ...671..990K, 2019A&A...622A.200K,2021A&A...647A.188D,2014Sci...346C.315P}, suggesting direct energy deposition. Quantitative analyses indicate that the energy released through network field cancellations, approximately 2 × 10$^5$ erg cm$^{-2}$ s$^{-1}$, closely matches the energy requirements for quiet corona heating \citep{1977ARA&A..15..363W}. Recent investigations have further highlighted the importance of small-scale magnetic interactions in driving larger-scale phenomena and contributing to the overall energy budget of the solar atmosphere \citep{2022ApJ...934...38L}.

A significant theoretical advance came with the development of the analytical cancellation nanoflare model by \citet{2018ApJ...862L..24P}. This model proposes that continuous photospheric motions drive widespread magnetic cancellation between opposite polarities, identifying photospheric converging flows as the fundamental trigger for chromospheric reconnection. The model provides a comprehensive framework connecting photospheric flows, chromospheric reconnection, photospheric flux cancellation, and atmospheric heating. These connections have been further supported by numerical simulations \citep{2019ApJ...872...32S, 2020ApJ...891...52S}, demonstrating how small-scale cancellation events can effectively channel energy into the solar atmosphere.

The cancellation nanoflare model is particularly significant for understanding quiet Sun heating. Unlike active regions, where large-scale magnetic structures dominate, quiet regions require a more distributed heating mechanism. The ubiquitous nature of small-scale cancellations, combined with their collective energy release, makes them promising candidates for maintaining coronal temperatures in these regions. However, despite the model's theoretical success and supporting simulations, the key features of small-scale cancellation events in observations are often covered by the chromospheric fibril canopy, making clear observational evidence for the nanoflare cancellation model less likely to be found.

This paper presents detailed observations of a small-scale cancellation nanoflare event, providing crucial observational support for the cancellation nanoflare model. The event, driven by chromospheric reconnection and featuring a current sheet approximately $2\arcsec$ in length, demonstrates the direct connection between photospheric flows and magnetic reconnection. Using multi-wavelength observations, we trace the complete evolution from initial photospheric flow patterns through magnetic field convergence to eventual reconnection and heating. Section 2 describes our observational data and methodology, Section 3 presents our detailed results, and Section 4 provides a comprehensive conclusion and discussion.

\section{Instruments and Data}
The present event was recorded simultaneously by space-based and ground-based telescopes. The Solar Dynamics Observatory \citep[SDO;][]{2012SoPh..275....3P} provided observations through two instruments: the Atmospheric Imaging Assembly \citep[AIA;][]{2012SoPh..275...17L} and the Helioseismic and Magnetic Imager \citep[HMI;][]{2012SoPh..275..229S}. AIA observed the coronal response at 171 \AA\ (characteristic temperature $\sim$10$^{5.8}$ K) with a 12-second cadence and $0\arcsec.6$ pixel size resolution. HMI measured photospheric line-of-sight magnetic fields with a 45-second cadence and $0\arcsec.5$ pixel size resolution. The Goode Solar Telescope \citep[GST;][]{2010AN....331..636C} at Big Bear Solar Observatory provided high-resolution chromospheric observations through its Visible Imaging Spectrometer (VIS), capturing H$\alpha$ wing ($\pm$0.8 \AA) images with 25-second cadence and $0\arcsec.029$ pixel size resolution.

All data were processed using standard reduction pipelines. The SDO data were aligned using the SolarSoft mapping routines. The GST data underwent speckle reconstruction and were coaligned with SDO observations using cross-correlation of common features. Differential emission measure analysis utilized six AIA EUV channels (94, 131, 171, 193, 211, and 335 \AA) to determine plasma temperatures and densities. The following equation determines the DEM:

\begin{equation}
	%\frac{\lambda}{\lambda_0}=\sqrt{\frac{C+V}{C-V}}
	{I_i}=\int{R_i}(T){\times}\rm{DEM}(\emph{T})d\emph{T},
\end{equation}
where ${I_i}$ is the observed emission intensity of a wavelength channel $i$; ${R_i}(T)$ is the temperature response function of the channel $i$; DEM($T$) represents the DEM of coronal plasma, computed through the method of I. G. Hannah and E. P. Kontar \citep{2012A&A...539A.146H}. After solving the DEM, one can calculate the Emission Measure (EM) within a temperature range ($T_{min}$, $T_{max}$) through the following equation:

\begin{equation}
	\rm{EM}=\mathit{\int_{T_{min}}^{T_{max}}}DEM(\emph{T})d\emph{T}.
\end{equation}	
Finally, one can estimate the average temperature through EM and DEM:
\begin{equation}
	\bar{T}=\frac{\int_{T_{min}}^{T_{max}}\mathrm{DEM}(T){\times}T\mathrm{d}T}{\rm{EM}},
\end{equation}
where $\bar{T}$ is the average temperature\citep{2012ApJ...761...62C}. The average electron number densities of the reconnection region $n_e$ is estimated according to $\sqrt{\rm{EM/H}}$ (assuming filling factor $\approx$ 1), where H is the depth along the line of sight (LOS). In this study, we assumed that the LOS depth equaled the height of the low corona bottom relative to the photosphere ($\sim$ 3 Mm).

\begin{figure*}
\epsscale{0.85}
\figurenum{1}
\plotone{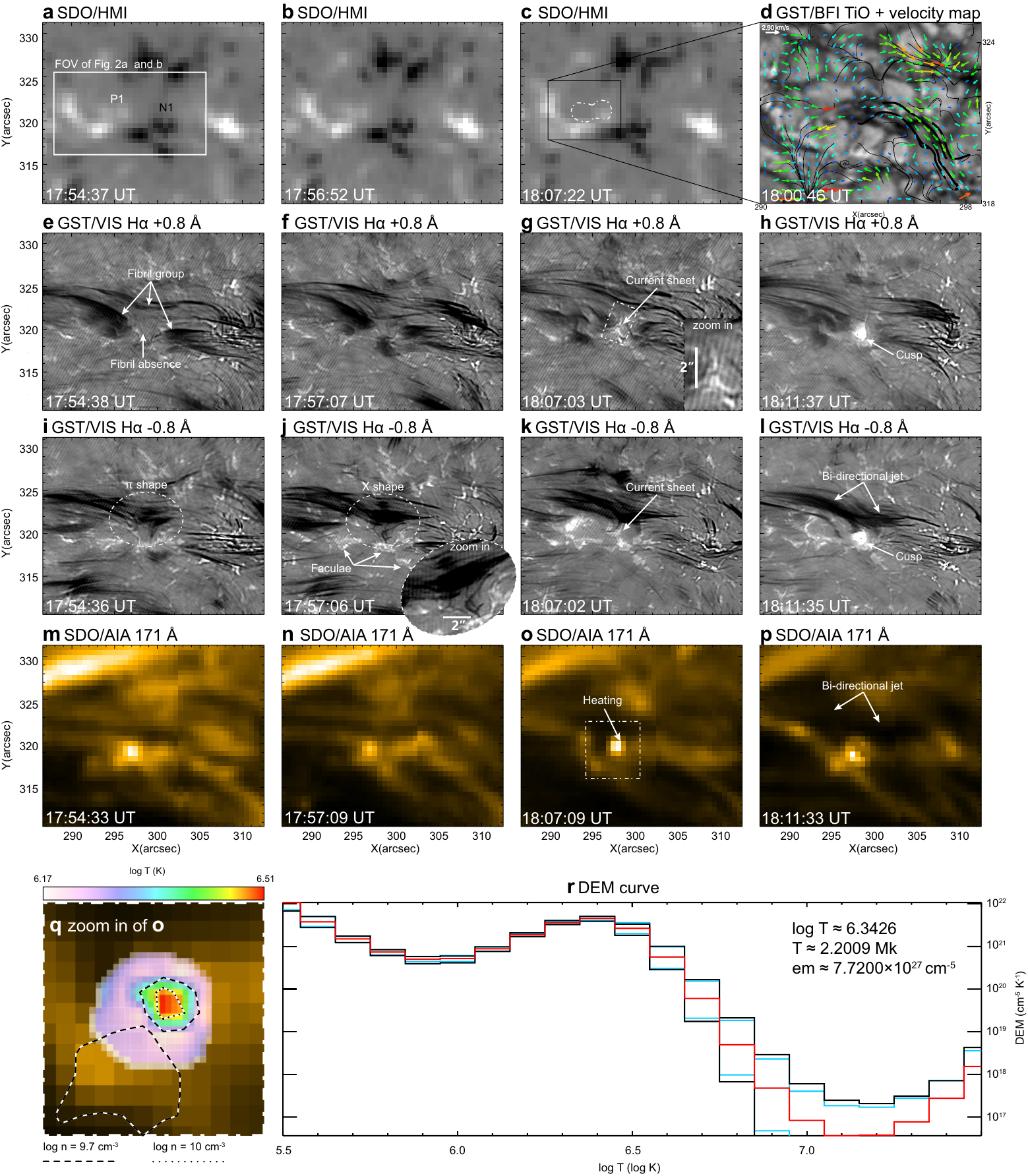}
\caption{Multi-wavelength evolution of the cancellation nanoflare from the photosphere to the corona. From top: SDO/HMI magnetograms, GST/VIS H$\alpha$ +0.8 \AA, H$\alpha$ -0.8 \AA, and SDO/AIA 171 \AA\ observations. The positive and negative magnetic fields are saturated at 500 G and -370 G, respectively. Initial magnetic polarities P1/N1 are marked in (a), with P1's starting position outlined in (c). The velocity map and streamlines in (d) show photospheric flows. Key features include fibril groups, $\pi$ and X-shaped configurations, current sheet formation, and coronal heating response. (q) the zoom-in of the white box region in (o). DEM analysis of brightening is overlaid on (o), which reveals log T(K) = 6.17--6.51 and log n = 9.7--10 cm$^{-3}$. (r) the DEM curve of the orange-red region in (q); the red line stands for the best-fitted DEM curve, while the black and blue curves represent the reconstructed curves from the 50 and 100 Monte Carlo (MC) simulations, respectively; see \citet{{2012ApJ...761...62C}} for details of MC simulations. An animation spanning 17:54--18:12 UT is available online, in which only the observations of the SDO/HMI, SDO/AIA 171 \AA, GST/VIS H$\alpha$ +0.8 \AA, H$\alpha$ -0.8 \AA, and GST/BFI TiO are included.
\label{fig1}}
\end{figure*}

\section{Results} 
\subsection{Layered Atmosphere Response}
We analyze a small-scale reconnection event on July 04, 2021, in a quiet region with approximately ($300\arcsec,  322\arcsec$) heliocentric coordinates. Figure 1 displays images of the region of interest (ROI) at different atmospheric layers, with a field of view of $21\arcsec \times 26\arcsec$. From the upper to bottom panels, we show the layered response of the ROI from the photosphere to the lower corona.

The first row shows SDO/HMI magnetograms, which map photospheric magnetic fields by measuring the Fe I Stokes profiles. In these magnetic field images (Fig. 1a--c), white and black areas represent positive and negative magnetic fields, respectively, revealing the evolution of photospheric magnetic fields. The second and third rows show GST/VIS H$\alpha$ +0.8 \AA\ (red wing) and H$\alpha$ -0.8 \AA\ (blue wing) images, respectively. The H$\alpha$ far-wing images reveal the photospheric continuum emission, appearing as a gray background and bright areas (faculae; see Fig. 1h). Comparing H$\alpha$ far-wing images with magnetic maps shows that photospheric faculae in GST/VIS H$\alpha$ far wings correspond to magnetic concentrations mapped in SDO/HMI. Magnetic concentrations enhance the photospheric intensity, which is why such strong correspondence exists. Moreover, H$\alpha$ far-wing images also show dark filaments. These dark filaments are absorption features formed by the radiative transfer process of the first spectral line of the Balmer series and represent chromospheric fibrils. Chromospheric fibrils most trace magnetic field lines \citep{2011A&A...527L...8D}. Thus, the evolution of chromospheric fibrils reveals how the magnetic connectivity of the ROI evolves both temporally and spatially. The last row shows SDO/AIA 171 \AA\ images. This channel is most sensitive to plasma at temperature $\sim$10$^{5.8}$ K. Due to resolution limitations, SDO/AIA cannot resolve the magnetic topology of the ROI as fine as GST/VIS. However, the EUV intensity evolution revealed by SDO/AIA 171 \AA\ observations shows the lower corona response of the ROI, which helps to analyze the heating process.

Our focus is on a pair of opposite polarities of ROI, which are respectively denoted by P1 and N1 in Fig. 1a. The time sequence runs left to right, showing pre-reconnection, null-point formation, current-sheet formation, and post-reconnection phases. As time progresses, positive polarity P1 slowly moves toward negative polarity N1 (Fig. 1a--c). P1's initial position is outlined in Fig. 1c for reference. From pre-reconnection (17:54:37 UT) to current-sheet formation (18:07:22 UT), P1 shifts approximately 3 Mm. Using the FLCT method \citep{2008ASPC..383..373F}, we calculate photospheric flow velocities and streamlines around P1, shown as arrows and black curves overlaid on the TiO image (Fig. 1d). The velocity map and streamlines reveal photospheric flows oriented from northeast to southwest. These flows appear as stretched dark lanes aligned with the streamlines in the TiO background, showing consistency with P1's migration toward N1 (see online animation).

\begin{figure*}
\epsscale{0.85}
\figurenum{2}
\plotone{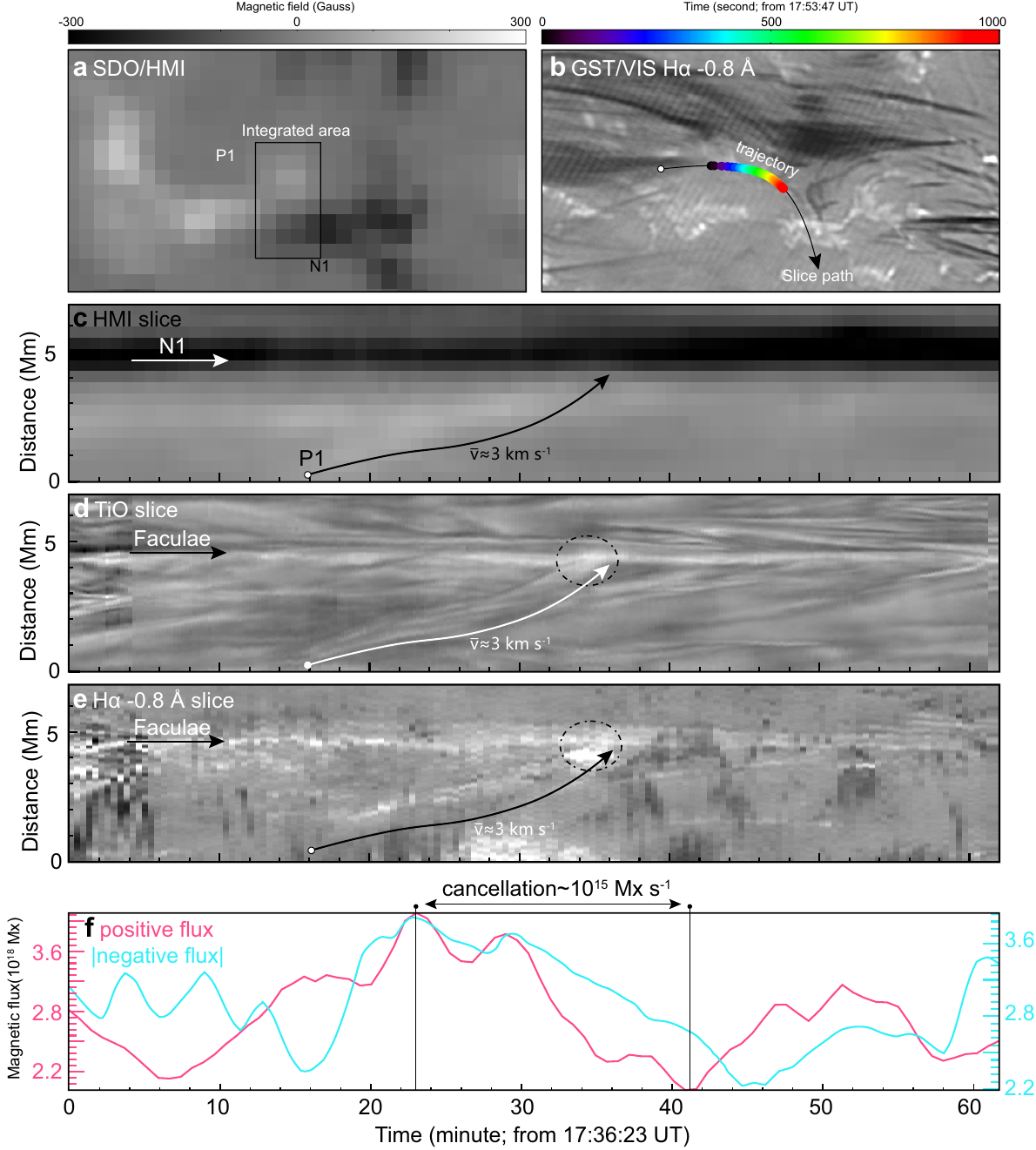}
\caption{Analysis of magnetic flux cancellation evolution. Panels (a-b) show the measurement region and slice path. Time-distance diagrams from HMI (c), TiO (d), and H$\alpha$ (e) reveal 3 km s$^{-1}$ convergence of opposite polarities. Dashed ellipses highlight interaction phases. Panel (f) tracks positive (red) and negative (blue) magnetic flux measured above $\pm$10 G, showing cancellation rate $\sim$10$^{15}$ Mx s$^{-1}$ during 17:59-18:17 UT. 
\label{fig2}}
\end{figure*}

The H$\alpha$ observations show two chromospheric fibril groups rooted in the magnetic polarity pair P1/N1. These fibril groups have different orientations: one group protrudes left while the other protrudes right (Fig. 1e, i). For convenience, we refer to these groups as left fibrils (LFs) and right fibrils (RFs). Another group of fibrils overlying the LFs and RFs exists, forming a $\pi$-shaped configuration in the H$\alpha$ blue wing during the pre-reconnection phase. Both H$\alpha$ wings show two key observational characteristics: 1) Magnetic field curvatures of LFs and RFs significantly decrease from footpoints N1 and P1 outward; 2) Fibrils are absent in the center of the fibril groups, as indicated by the arrow in Fig. 2e. These characteristics respectively suggest significant magnetic gradients around the central area and weak magnetic field strength in the fibril-absent region.

During null-point formation, as P1 moves closer to N1, the LFs approach the RFs. Consequently, the area of fibril absence becomes smaller, and the curvatures of adjacent fibrils increase. A small-scale fibril ($<2\arcsec$) appears below the $\pi$-shaped region (Fig. 1j zoom-in), resulting from separatrix rise between fibril groups. The small-scale fibril and $\pi$-shaped fibril group together to form an X-shaped configuration.

When the magnetic system enters current-sheet formation, P1 approaches N1 further, causing LFs and RFs to collide. A bright elongated structure (~$2\arcsec$) forms where fibrils were previously absent (Fig. 1g,k). Movies show fibrils flowing inward from both sides, breaking and reconnecting into roughly horizontal outflow fibrils \citep{2019ApJ...883..104S, 2024ApJ...974L...3Z}, similar to observations of larger-scale reconnection events \citep{2019ApJ...885L..11S}. Post-reconnection, a cusp-shaped loop forms downstream (Fig. 1h,l) with bi-directional jets visible in the blue wing. This small-scale cusp shares similar magnetic topology with large-scale reconnection events \citep{2011ApJ...742...82K, 2016NatPh..12..847L, 2008ApJ...683L..83N, 2019ApJ...883..104S}, suggesting topological similarity between small and large-scale reconnection systems.

Based on H$\alpha$ observations, we identify the bright elongated structure as a current sheet - a dissipation region favoring frozen-flux violation and field reconnection \citep{Dungey01071953}. The X-shaped region represents a magnetic null point, a separator layer enabling current sheet formation \citep{2022LRSP...19....1P,2014masu.book.....P}. These features form through continuously increasing stress between converging fibril groups, consistent with earlier observations of significant magnetic gradients and weak fields near the fibril center.

During current sheet formation, comparing H$\alpha$ wing images (Fig. 1g,k) with 171 \AA\ observations (Fig. 1o) reveals coronal brightening at exactly the same location, indicating localized heating. DEM result \citep{2012ApJ...761...62C} shows that the significant heating area, characterized by the orange-red area in Fig. 1q, is approximately 1 Mm$^2$, whose electron densities and plasma averaged temperature reaching $\sim$ 10$^{10}$ cm$^{-3}$ (Fig. 1q) and $\sim$ 2.2 Mk (Fig. 1q and r), respectively. The bi-directional jet later appears in 171 \AA\ emission (Fig. 1p), emanating from the heated region \citep{2019ApJ...883..104S}.

\begin{figure*}
\epsscale{0.85}
\figurenum{3}
\plotone{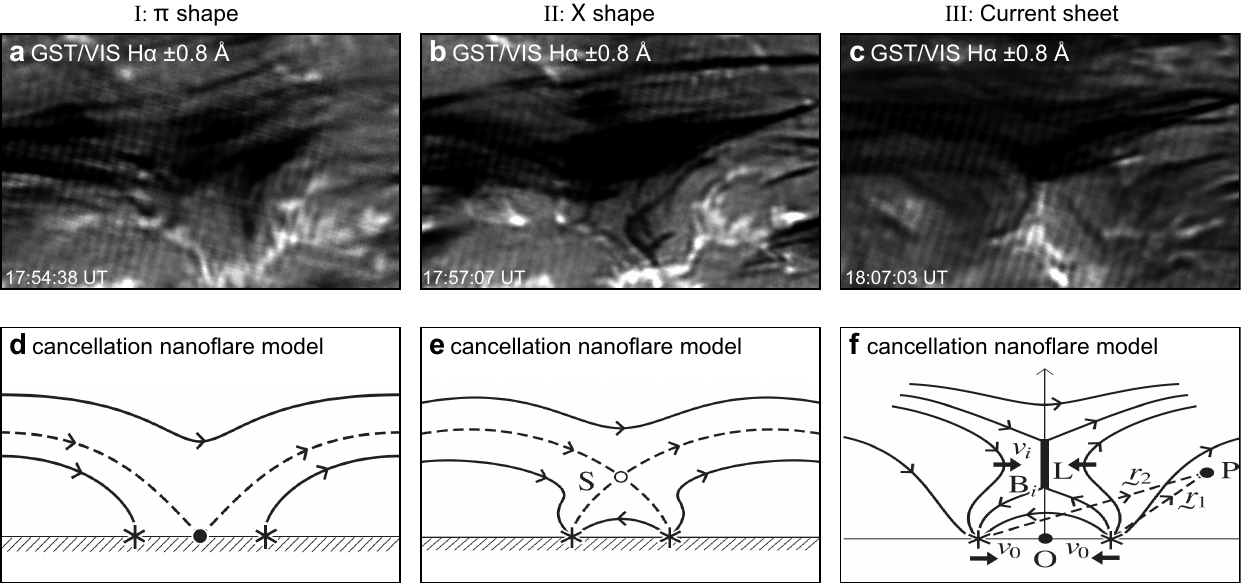}
\caption{Comparison between observations and cancellation nanoflare model predictions. (a -- c) GST/VIS H$\alpha$ observations showing evolution through $\pi$-shaped, X-shaped, and current sheet phases. (d--f) Corresponding magnetic field configurations from \citet{2018ApJ...862L..24P}, with dashed lines indicating magnetic separatrices. An animation of the evolutionary sequence is available.
\label{fig3}}
\end{figure*}

\subsection{The Trigger and Photospheric Footprint of the Reconnection}
The layered atmospheric observations (Fig. 1) reveal tight spatial-temporal correlations between photospheric magnetic polarity convergence, chromospheric fibril approach, current sheet formation, and coronal brightening. To analyze these relationships in detail, we create temporal-spatial slices from SDO/HMI, GST/BFI TiO, and GST/VIS H$\alpha$ along the path shown in Fig. 2b. These three slices respectively demonstrate the temporal evolution of photospheric magnetic polarities (Fig. 2c), photospheric granules (Fig. 2d), and photospheric faculae (Fig. 2e) along the same spatial path.

In the HMI slice, the polarity inversion line (PIL) lies near y = 4 Mm, with P1 and N1 residing above and below the PIL, respectively. This slice reveals three significant features. First, the magnetic polarity N1 shows minimal movement throughout the slice, remaining almost stationary at y = 5 Mm. Second, at 17:52 UT, positive polarity P1 begins moving toward N1 at an average speed of about 3 km s$^{-1}$, eventually colliding with N1 at 18:12 UT. The third notable feature is that closely following the collision, the positive flux shows significant weakening.

Due to their higher resolution, the TiO and H$\alpha$ slices reveal more detailed evolution. In both slices, the unmoving polarity N1 corresponds to a stationary photospheric facula (see Fig. 2d), while the moving positive polarity P1 corresponds to a photospheric facula moving at an averaged speed of approximately 3 km s$^{-1}$. The TiO slice also reveals that the slope of the temporal-spatial trajectory of this photospheric facula was temporarily reduced from 17:58 UT to 18:03 UT, suggesting its speed slowed down during this period. Regardless of motion dynamics, the temporal-spatial trajectory of magnetic polarities displays strong consistency with that of photospheric faculae. Photospheric bright points form through magnetic concentration, appearing in the continuum spectrum as faculae. In other words, photospheric faculae and magnetic polarities are simply different manifestations of the same phenomenon, explaining their strong temporal-spatial correlation.

\begin{table*}[htbp]
%\centering
\caption{Parameter Comparison between the Observation and the Cancellation Nanoflare Model}
\hspace*{-1.5cm}
%\renewcommand\arraystretch{0.5}
%\begin{tabular}{|l|l|l|l|l|l|l|}
\resizebox{\linewidth}{!}{
\begin{tabular}{|c|c|c|c|c|c|c|c|c|}
\hline
                           & {\makecell*[c]{\hspace{-3em}heating\\ \hspace{-3em}area ($S$)}}  & {\makecell*[c]{\hspace{-3em}magnetic field  \\  \hspace{-3em}strength ($B$)}}  & {\makecell[c]{\hspace{-3em}total flux source\\   \hspace{-3em}speed ($v$)}} & {\makecell*[c]{\hspace{-3em}cancellation \\  \hspace{-3em}flux ($\phi$)}} & {\makecell*[c]{\hspace{-3em}cancellation\\ \hspace{-3em}duration ($t$)}} & {\makecell*[c]{\hspace{-3em}cancellation\\ \hspace{-3em}rate ($\frac{\phi}{t}$)}} &  {\makecell[c]{\hspace{-3em}energy conversion\\  \hspace{-3em}rate ($\frac{E}{t}$)}} &{\makecell*[c]{ \hspace{-3em}heating\\ \hspace{-3em}rate ($\frac{E}{St}$)}} \\ 

\hline
observation                              &       10$^{16}$ cm$^2$     & 20 G     & 3 km s$^{-1}$           & 1.1$\times$10$^{18}$ Mx & 1080 s &  10$^{15}$ Mx s$^{-1}$  &     8.7$\times$10$^{22}$ erg s$^{-1}$  &    8.7$\times$10$^6$ erg cm$^{-2}$ s$^{-1}$           \\ 

\hline
 {\makecell[c]{\hspace{-3em}cancellation nanoflare\\  \hspace{-3em}model}} &   10$^{16}$ cm$^2$    & 10 G                                   & 2 km s$^{-1}$              & 10$^{18}$ Mx & 1000 s &  10$^{15}$ Mx s$^{-1}$   &    5$\times$10$^{22}$ erg s$^{-1}$    &    5$\times$10$^6$ erg cm$^{-2}$ s$^{-1}$          \\ \hline
\end{tabular}

}
\end{table*}

The convergence and collision process appears more clearly in the TiO and H$\alpha$ slices. As shown in Fig. 2d and e, the two faculae began converging at 17:52 UT and eventually collided at 18:12 UT. To study the magnetic flux evolution quantitatively during the collision between N1 and P1, we measure the magnetic flux within an area surrounding the polarity inversion line of P1 and N1 (see Fig. 2a). The measured positive and absolute negative magnetic fluxes are shown in Fig. 2f. During the collision between magnetic polarities N1 and P1 or two photospheric faculae, the positive and negative fluxes show significant mutual cancellation, consistent with recent high-resolution observations \citep{2018ApJ...861..105S, 2023ApJ...942L..22D}. Within about 1080 seconds (from 17:59 UT to 18:17 UT), the canceled positive flux is about 1.8×10$^{18}$ Mx, while the canceled negative flux is about 1.1×10$^{18}$ Mx. The canceled negative flux represents the lower bound of the canceled value (~1.1×10$^{18}$ Mx). Dividing this minimum canceled flux by the cancellation period yields a cancellation rate of 10$^{15}$ Mx s$^{-1}$ at least.

The occurrence of magnetic cancellation (17:59 UT -- 18:17 UT) aligns closely with the collision and convergence between P1 and N1 (17:52 UT -- 18:12 UT), indicating a strong temporal and spatial correlation between magnetic cancellation and convergence of opposite magnetic polarities. Such a close relationship suggests that the magnetic convergence of opposite magnetic polarities facilitates their magnetic cancellation, as suggested by \citet{1985AuJPh..38..929M}. While the converging motion of magnetic fields can be considered the immediate cause facilitating the magnetic cancellation, it is not the fundamental cause. Since magnetic fields are frozen into high-$\beta$ photospheric flows, they are forced to move under the stress of photospheric flows, which are revealed in Fig. 2d and e. Thus, the convergence of magnetic fields is essentially driven by high-$\beta$ photospheric flows. These flows are the underlying drivers behind both magnetic convergence and cancellation. As photospheric flows continuously stress opposite magnetic polarities and cause their convergence, a current sheet will eventually be formed between chromospheric fibrils rooted in these stressed polarities, as seen in Fig. 1g and k.

\subsection{Comparison with the Cancellation Nanoflare Model}
The topological evolution of this event shows remarkable similarity with the cancellation nanoflare model of \citet{2018ApJ...862L..24P}. Figure 3 highlights this correspondence. Panels Fig. 3a--c show zoomed views from Fig. 1d+g, 1e+h, and 1f+i, emphasizing the temporal evolution of magnetic topology during cancellation. The chromospheric fibrils exhibit three distinct configurations: $\pi$-shaped (Fig. 3a), X-shaped (Fig. 3b), and current sheet (Fig. 3c).

The analytical predictions from \citet{2018ApJ...862L..24P} are shown in Fig. 3d--f for direct comparison. These panels depict the model's predicted magnetic topology evolution from initial configuration through the approaching phase to current sheet formation. A side-by-side comparison reveals striking similarities between observations (Fig. 3a--c) and theoretical predictions (Fig. 3d--f).

We quantitatively estimate key physical parameters, summarized in Table 1. The heating area $S$ is estimated by the orange-red region in Fig. 1q, while the magnetic field strength $B$, total flux source speed $v$, cancellation flux $\phi$, cancellation duration $t$, and cancellation rate $\phi/t$ are directly measured from observations in Fig. 2. The energy conversion rate and heating rate are estimated as $E/t$ and $E/St$ respectively ($E = \frac{1}{8\pi}B\phi L$; $L$ is the current sheet length). Corresponding parameters from \citet{2018ApJ...862L..24P} are included in Table 1 for comparison. Beyond topological similarities, both observations and models share comparable physical parameters. This agreement in morphology and quantitative measures provides robust observational support for the small-scale cancellation nanoflare model.

\section{Conclusions and Discussions}
This study provides compelling observational evidence for magnetic cancellation as a fundamental mechanism for solar atmospheric heating. Using high-resolution multi-wavelength observations from GST and SDO, we present a detailed analysis of a small-scale magnetic cancellation event that demonstrates the direct connection between photospheric dynamics and atmospheric heating through magnetic reconnection \citep{2019A&A...622A.200K, 2018ApJ...864...68S}.

The observed magnetic topology evolution shows remarkable correspondence with theoretical predictions of the cancellation nanoflare model \citep{2018ApJ...862L..24P}. We trace the complete sequence from the initial $\pi$-shaped configuration through the X-shaped topology to the current sheet formation, providing direct observational validation of the model's predicted evolution. This correspondence extends beyond morphology to quantitative parameters - our measured cancellation rate ($\sim$10$^{15}$ Mx s$^{-1}$), heating area ($\sim$10$^{16}$ cm$^2$), and magnetic field strength ($\sim$20 G) closely match model expectations. Most significantly, the derived heating rate of 8.7$\times$10$^6$ erg cm$^{-2}$ s$^{-1}$ approaches that required for sustaining coronal temperatures \citep{1977ARA&A..15..363W}.

The high-resolution observations reveal how photospheric flows drive the convergence of opposite-polarity fields, leading to magnetic stress accumulation and eventual reconnection in the chromosphere. This mechanism provides crucial insight into the surface-corona coupling process, demonstrating how small-scale surface dynamics can influence larger atmospheric volumes through magnetic connectivity \citep{2018ApJ...861..105S, 2017ApJ...851...67S}. The observed heating signatures in multiple AIA channels confirm effective energy transport from the photosphere to the corona through this process.
Recent statistical studies have shown that small-scale magnetic cancellations occur far more frequently than previously thought \citep{2017ApJS..229...17S}, with a combined cancellation rate reaching 10$^{15}$ Mx s$^{-1}$. Our detailed observations of a single event demonstrate that these cancellations can indeed drive significant atmospheric heating through reconnection, supporting their role as a substantial contributor to the coronal energy budget \citep{2019ApJ...872...32S, 2020ApJ...891...52S}.

The multi-scale nature of the heating process is particularly noteworthy. While the initial cancellation occurs over $\sim$1 Mm at the photosphere, the induced chromospheric reconnection and subsequent heating extend several Mm into the atmosphere. This illustrates how relatively small-scale photospheric dynamics can affect larger coronal volumes through magnetic coupling \citep{2018ApJ...861..105S}. The heating signature observed in multiple AIA channels, along with derived mega-Kelvin temperatures, confirms the effectiveness of this energy transport mechanism.

Another important point to consider is the global significance of these small-scale events for the quiet Sun heating. The statistical result in \citet{2024A&A...683A.190R} suggested that the number of small-scale reconnections $N$ is more than 750,000 at any time in the deep atmosphere of the quiet Sun. Since most of their statistical samples occur at weak magnetic environments similar to the one reported here, we simply assume that the energy conversion rate of the small-scale reconnections in \citet{2024A&A...683A.190R} is comparable to ours ($\sim$ 8.7$\times$10$^{22}$ erg s$^{-1}$). Based on such simple assumption, the global heating rate of these small-scale reconnections is $\frac{NE}{S_\odot{t}}\approx$ 1.1$\times$10$^6$ erg cm$^{-2}$ s$^{-1}$, where $S_\odot$ represents the area of the low solar atmosphere ($\sim$ 6.1$\times$10$^{22}$ cm$^2$). This estimated value matches the energy flux requirements for the quiet Sun heating \citep{1976ASSL...53.....A,1977ARA&A..15..363W}. However, it is not time to conclude that this small-scale reconnection can heat the quiet Sun due to the lack of rigorous examinations. It merely means the possibility of these small-scale reconnections heating the quiet Sun. Further statistical studies will be particularly important for more accurately quantifying the global significance of cancellation events in solar atmospheric energetics.

Our results demonstrate that magnetic cancellation at the solar surface can drive significant atmospheric heating through reconnection, supporting the cancellation nanoflare model as a viable mechanism for solar atmospheric heating. The combination of high-resolution observations and theoretical understanding suggests these small-scale events may be fundamental building blocks in the complex coronal heating process \citep{ 2024ApJ...960...51P}. The striking correspondence between observed features and model predictions, from magnetic topology evolution to quantitative parameters, strongly validates this heating mechanism.

The relationship between small-scale magnetic cancellation events and larger solar phenomena reveals fascinating scale-dependent behavior while maintaining some fundamental similarities. The small-scale event described in this paper ($\thicksim$1 Mm) represents the basic building blocks of magnetic energy release in the solar atmosphere. Several key differences emerge when compared to solar jets ($\thicksim$5--10 Mm). Jet-producing cancellation events typically involve stronger magnetic fields (100--500 G versus $\thicksim$20 G in our case) and higher cancellation rates ($10^{16}-10^{17}$ Mx/s versus $10^{15}$ Mx/s) \citep{2007A&A...469..331J, 2010ApJ...720..757M, 2012ApJ...745..164S, 2017ApJ...851...67S, 2018ApJ...864...68S, 2021RSPSA.47700217S}. Jets and magnetohydrodynamic waves often involve mini-filament eruptions as part of their formation mechanism, adding complexity to the magnetic topology \citep{2012ApJ...745..164S, 2019ApJ...883..104S, 2021RSPSA.47700217S, 2022ApJ...926L..39D,2014ApJ...786..151S,2014ApJ...795..130S}. At even larger scales, filament eruptions (tens to hundreds of Mm) showcase how sustained magnetic cancellation along polarity inversion lines can gradually build up twisted magnetic flux ropes \citep{1989ApJ...343..971V, 2021ApJ...923...45Z, 2024ApJ...964..125S}. These large-scale events typically have the strongest fields and highest cancellation rates among the three categories. However, despite these scale-dependent differences, all three phenomena share a common underlying process: the convergence of opposite-polarity magnetic fields leading to reconnection through current sheet formation. This suggests that magnetic cancellation may be a fundamental mechanism operating across different scales in the solar atmosphere. Small-scale events like the one in this study could act as elementary energy release events that sometimes combine and build up to drive larger eruptions. This multi-scale perspective has important implications for understanding how magnetic energy is transported from the photosphere into the corona. Small-scale cancellation events may be the initial links in an energy transport chain that ultimately powers larger solar eruptions. Future statistical studies comparing cancellation events across scales could help clarify these connections and their role in solar atmospheric heating and dynamics.

\begin{acknowledgments}
The authors thank the SDO and the GST teams for providing excellent data and the anonymous referee for providing many valuable comments to improve the quality of this Letter. This work is supported by the Natural Science Foundation of China (12173083) and the Shenzhen Key Laboratory Launching Project (No. ZDSYS20210702140800001) and  the Specialized Research Fund for State Key Laboratory of Solar Activity and Space Weather.
\end{acknowledgments}

%\bibliographystyle{aasjournal}
%\bibliography{bibfile}

\end{document}